\documentclass[aps,prd,twocolumn,groupedaddress,showpacs,showkeys,nofootinbib]{revtex4}%
\usepackage{mathrsfs} 
\usepackage{graphicx}
\usepackage{bm}
\usepackage{amssymb,amsmath,amsthm}
\begin{document}%
\title{\bf
Power-counting renormalizability of generalized Ho\v{r}ava gravity
}
\author{Matt Visser}
\affiliation{School of Mathematics, Statistics, and Operations Research,
Victoria University of Wellington, New Zealand}
\date{24 December 2009;
\LaTeX-ed \today}
\begin{abstract}
In an earlier article [arXiv: 0902.0590 [hep-th], Phys.\ Rev.\ {\bf D80} (2009) 025011], I discussed the potential benefits of allowing Lorentz symmetry breaking in quantum field theories. In particular I discussed the perturbative power-counting \emph{finiteness} of the normal-ordered $:\!P(\phi)^{z\geq d}_{d+1}\!:$ scalar quantum field theories, and sketched the implications for Ho\v{r}ava's model of quantum gravity. In the current rather brief addendum, I will tidy up some dangling issues and fill out some of the technical details of the argument indicating the power-counting \emph{renormalizability} of a $z\geq d$ variant of Ho\v{r}ava gravity in $(d+1)$ dimensions.
\end{abstract}
\maketitle
\def\lint{\hbox{\Large $\displaystyle\int$}} 
\def\hint{\hbox{\Huge $\displaystyle\int$}}  

\def\d{{\mathrm{d}}}
\newcommand{\scri}{\mathscr{I}}
\newcommand{\sun}{\ensuremath{\odot}}

\section{Introduction}

In reference~\cite{LSB-horava} I argued for the perturbative power-counting \emph{finiteness} of the normal-ordered $:\!\!P(\phi)^{z= d}_{d+1}\!\!:$ scalar quantum field theories in $(d+1)$ dimensions, where the central defining feature of these quantum field theories is that the kinetic term in the Lagrangian contains exactly two time derivatives and up to $2d=2z$ space derivatives.

 I then rather briefly sketched the implications of this result for the perturbative power-counting \emph{renormalizability} of a suitably defined $(d+1)$ dimensional version of Ho\v{r}ava gravity, where the central defining feature of this model is again the presence of exactly two time derivatives and up to to $2d=2z$ space derivatives in the Lagrangian. 
This is a natural generalization of the specific  $z=3$ modification of $(3+1)$ gravity that was explicitly introduced by Ho\v{r}ava~\cite{Horava}.  

In that earlier article~\cite{LSB-horava}, I terminated the discussion at the stage where it became clear that the perturbatively quantized graviton  field could consistently be assigned a canonical dimension of zero --- this being the standard perturbative signal that arbitrarily high-order Feynman diagrams behave no worse than low-order Feynman diagrams, thereby \emph{implicitly} implying (power-counting) renormalizability. However I did not include any \emph{explicit} power-counting argument. Since this omission has led to some ongoing confusion, I will in this brief addendum to the previous article~\cite{LSB-horava}  tidy up the argument by providing the missing details.

 In particular, I will supplement that previous discussion with a more explicit argument involving the superficial degree of divergence, and will also take advantage of this opportunity to extend the discussion to consider the situation for $z>d$. Despite the considerable activity regarding  other aspects of Ho\v{r}ava gravity, regarding renormalizability so far very little has been done that goes beyond simple power counting arguments. 
 
 \vfill

\section{Power-counting}

Recall that for the  $P(\phi)^{z}_{d+1}$ scalar quantum field theories each loop integral has dimension $ [\kappa]^{d+z}$, while each propagator has dimension $ [\kappa]^{-2z}$. To analyze the superficial degree of divergence one need only consider the one-particle-irreducible (1PI) sub-diagrams of the  Feynman diagram. For each such 1PI  sub-diagram the total contribution to dimensionality coming from loop integrals and internal propagators is $[\kappa]^{(d+z)L-2Iz}$,
which is summarized by saying that  the ``superficial degree of divergence'' is
\begin{equation}
\delta = (d+z)L-2Iz = (d-z)L-2(I-L)z.
\end{equation}
Note that the quantity $I$ only counts the propagators internal to  the 1PI sub-diagram.
But to get $L$ loops one needs, at the very least, $I$ internal propagators. So for any 1PI Feynman diagram we certainly have
\begin{equation}
\delta \leq  (d-z)L.
\end{equation}
Consequently, if one picks $z\geq d$ then
\begin{equation}
\delta \leq  0,
\end{equation}
and the \emph{worst} divergence one can possibly encounter is logarithmic. (Or a power of a logarithm if one has several subgraphs with $\delta=0$.) 

Since at this stage of the argument we have already assumed  $z\geq d$, such a logarithmic divergence can occur only for the borderline case $z=d$ and $L=I$, that is for a ``rosette'' Feynman diagram.  This observation is enough to guarantee that the non-normal-ordered  $P(\phi)^{z=d}_{d+1}$ is power-counting \emph{renormalizable}, and to render the normal-ordered   $:\!P(\phi)^{z=d}_{d+1}\!:$ power-counting \emph{finite}. Furthermore if one takes $z>d$ this discussion is sufficient to render $P(\phi)^{z>d}_{d+1}$ (with or without normal ordering) power-counting \emph{finite}.

\clearpage

Turning our attention now to  a $z\geq d$ variant of Ho\v{r}ava gravity in $(d+1)$ dimensions, (containing up to $2z$ spatial derivatives of the $d$ dimensional spatial metric), one obtains the same power-counting for the loop integrals and the propagators --- \emph{the difference now lies in the graviton self-interaction vertices}. While the vertices for the scalar field theory carried no factors of momentum, for Ho\v{r}ava gravity and its variants the graviton self-interaction vertices arise from a perturbative action of the form~\cite{LSB-horava}
\begin{equation}
S \sim \int \left\{ \dot h^2 + P(\nabla^{2z},h)\right\} \;\d t \, \d^d x,
\end{equation}
where $P(\nabla^{2z},h)$  is now an infinite-order polynomial in the graviton field $h$, which contains up to $2z$ spatial derivatives.  

In contrast to the scalar self-interaction vertices, the graviton self-interaction vertices thus contain up to $2z$ factors of momentum. 
If these are external momenta they do not contribute to the superficial degree of divergence. However internal momenta, and for any 1PI  Feynman diagram with $V$ vertices there can be up to $2z V$ factors of internal momenta, do contribute to the superficial degree of divergence. Consequently we now have the inequality
\begin{equation}
\delta \leq (d+z)L+ 2z(V-I) = (d-z)L+2z(V+L-I).
\end{equation}
But as always, Euler's theorem for graphs implies 
\begin{equation}
V+L-I=1
\end{equation}
so that
\begin{equation}
\delta \leq   (d-z)L+2z.
\end{equation}
For $z \geq d$ one simply has
\begin{equation}
\delta \leq  2z.
\end{equation}
Thus the superficial degree of divergence of the 1PI Feynman diagrams is  bounded by the canonical dimension of the operators already explicitly included in the bare action. This is the standard signal for renormalizability. 
As always, one should include in the bare action all terms compatible with the power counting and the symmetries of the theory. But that is exactly what the $z\geq d$ variant of Ho\v{r}ava gravity in $(d+1)$ dimensions is designed to do, and we conclude that any $z\geq d$ variant of Ho\v{r}ava gravity in $(d+1)$ dimensions is power-counting renormalizable. 
Note that instead of stopping at $z=d$ as is usually done, the present argument applies to all $z\geq d$. 

\section{Discussion}

In the specific case of $(3+1)$ dimensions, the minimal condition to get renormalizability is $z=3$. This is the situation that is most commonly considered. The situation where ``detailed balance''  is invoked to suppress some terms is most clearly and forcefully introduced  in~\cite{Horava}, with additional detail provided in~\cite{criticality, spectral}. In contrast, if one abandons detailed balance then one should include all possible terms up to $z=3$, as has forcefully been advocated in~\cite{nine, nine2}.

If one wishes to go beyond power-counting for Ho\v{r}ava gravity then the renormalizability arguments are as yet woefully incomplete. In the absence of gravity, some very useful indicative results are those of Anselmi and Halat~\cite{Anselmi1}, and Anselmi~\cite{Anselmi2, Anselmi3, Anselmi4, Anselmi5},  where the perturbative renormalizability of $z=d$ scalar--fermion--Yang--Mills field theories in flat space\-time have been investigated in considerable detail. Some progress using stochastic quantization (but limited to situations in which detailed balance applies) is reported by Orlando and Reffert in reference~\cite{orlando}. Extensions of this idea are reported in reference~\cite{orlando2}. (See also Shu and Wu in reference~\cite{shu}.)  Explicit renormalization group calculations are reported by Iengo, Russo, and Serone in reference~\cite{iengo}. (See also Collins \emph{et al.}~\cite{collins} for a more general analysis indicating  the generic necessity for fine tuning in Lorentz violating theories.) More recently, Alexandre \emph{et al.}~\cite{dyson} have used non-perturbative  Schwinger--Dyson techniques to investigate $z=3$ (fermion+scalar) Yukawa field theories  in flat (3+1) dimensional spacetime. In counterpoint, in the original articles~\cite{Horava, criticality, spectral} Ho\v{r}ava several times referred to his model as ``arguably finite'' --- to the best of my knowledge no significant  progress along those lines has been made.
Because the current argument is limited to power counting, it has nothing specific to say, pro or con, regarding the vexatious issue of the scalar graviton.

\section*{Acknowledgments:} I wish to thank Jianwei Mei for some penetrating questions that encouraged me to revisit this question in more detail.
This research was supported by the Marsden Fund administered by the Royal Society of New Zealand. Parts of this addendum were inspired by numerous discussions at the ``Emergent Gravity IV'' conference.



\end{document}